\documentclass[conference]{IEEEtran}
\IEEEoverridecommandlockouts

\makeatletter
\providecommand\sf@counterlist{}
\makeatother

\usepackage{cite}
\usepackage{amsmath,amssymb,amsfonts}
\usepackage{algorithmic}
\usepackage{graphicx}   
\usepackage{caption}    
\usepackage{subcaption} 
\usepackage{subfig}   
\usepackage{eurosym}
\usepackage{siunitx}
\usepackage{multirow}

\usepackage{lmodern}

\DeclareUnicodeCharacter{03A6}{\ensuremath{\Phi}}

\usepackage[T1]{fontenc}
\usepackage[utf8]{inputenc}
\usepackage{authblk}
\usepackage{newtxtext,newtxmath}

\usepackage{orcidlink}
\usepackage{academicons}
\definecolor{orcidlogocol}{HTML}{A6CE39}

\usepackage{textcomp}
\usepackage{xcolor}
\definecolor{good}{RGB}{200,255,200}  
\definecolor{bad}{RGB}{255,210,210}   

\usepackage[font=footnotesize,labelfont=bf]{caption}

\usepackage{tabularx}
\usepackage{dblfloatfix} 
\usepackage{booktabs}
\usepackage{paralist}
\usepackage{enumitem}
\usepackage[export]{adjustbox}
\usepackage{multicol}
\usepackage{gensymb}
\usepackage{steinmetz}
\usepackage{mathtools}
\usepackage{siunitx}
\usepackage{wrapfig}
\usepackage{subfig}
\usepackage[english]{babel}
\usepackage{blindtext}
\usepackage{setspace}
\usepackage{cancel} 

\usepackage{tikz}   

\usepackage[nolist,nohyperlinks]{acronym}

\usepackage[absolute,overlay]{textpos}

\newcommand{\copyrightbar}{%
  \setlength{\fboxrule}{.4pt}   
  \setlength{\fboxsep}{4pt}     
  \fcolorbox{black}{gray!15}{%
    \parbox{\dimexpr\textwidth-1\fboxsep-2\fboxrule}{%
      \centering\scriptsize
      CC-BY-NC-SA. Preprint of "\href{https://ieeexplore.ieee.org/author/739379890705013}{A Data-Based Review of Battery Electric Vehicle and Traction Inverter Trends}", accepted for 2025 IECON, 51st Annual Conference of the IEEE IES.\\[3pt]%
    }%
  }%
}

\def\BibTeX{{\rm B\kern-.05em{\sc i\kern-.025em b}\kern-.08em
    T\kern-.1667em\lower.7ex\hbox{E}\kern-.125emX}}

\let\OLDthebibliography\thebibliography
\renewcommand\thebibliography[1]{
  \OLDthebibliography{#1}
  \setlength{\parskip}{0pt}
  \setlength{\itemsep}{0pt plus 0.3ex}
}

\hypersetup{hidelinks}

\begin{document}

\captionsetup[figure]{labelfont={bf},name={Fig.},labelsep=colon}
\captionsetup[table]{labelfont={bf},name={TABLE},labelsep=colon}

\title{A Data-Based Review of Battery Electric Vehicle and Traction Inverter Trends\\
}






\author[$\ddagger$,$\S$,\orcidlink{0009-0005-5559-9342}]{Christoph~Sachs}
\author[$\S$]{Martin~Neuburger}

\affil[$\ddagger$]{\textit{Institute for System Dynamics, University of Stuttgart, Germany}}
\affil[$\S$]{\textit{Faculty of Mobility and Technology, Esslingen University, Germany}}
\affil[ ]{\orcidlink{0009-0005-5559-9342}~0009-0005-5559-9342}



\maketitle

\begin{textblock*}{\textwidth}(40pt,730pt)  
\copyrightbar
\end{textblock*}

\begin{abstract}
\acp{BEV} have advanced significantly during the past decade, yet drivetrain energy losses continue to restrict practical range and elevate cost. A dataset comprising more than 1\,000 European‑market \acp{BEV} (model years 2010--2025) is combined with detailed inverter–motor co‑simulation to chart technology progress for and quantify the efficiency and cost‑saving potential of partial‑load optimised \ac{MLI} for 2030. Average drive‑cycle range has climbed from 135\,km to 455\,km, while fleet‑average energy consumption has remained virtually constant. 

Three inverter topologies are assessed to evaluate future efficiency and cost enhancements: a conventional \ac{2L} \ac{B6} inverter with \ac{Si} and \ac{SiC} devices, and two \ac{3L} \ac{TNPC} and \ac{ANPC} inverters tailored for partial‑load operation. The \ac{3L}-\ac{TNPC} inverter, realised with only 30\,\% additional \ac{SiC} chip area, lowers drive‑cycle drivetrain losses by 0.67\,${}^{\text{kWh}}\!/\!_{\text{100\,km}}$ relative to a \ac{SiC} \ac{2L}-\ac{B6} baseline. These results identify partial‑load optimised \acp{MLI} as a cost‑effective route to further reduce \ac{BEV} energy consumption and total system cost.

\end{abstract}

\begin{IEEEkeywords}
Battery electric vehicles, traction inverter, Multilevel converter, silicon carbide, drivetrain efficiency, cost analysis
\end{IEEEkeywords}



\begin{figure}[b!]
\vspace{-0.5cm}
    \centering
    \includegraphics[width=\linewidth]{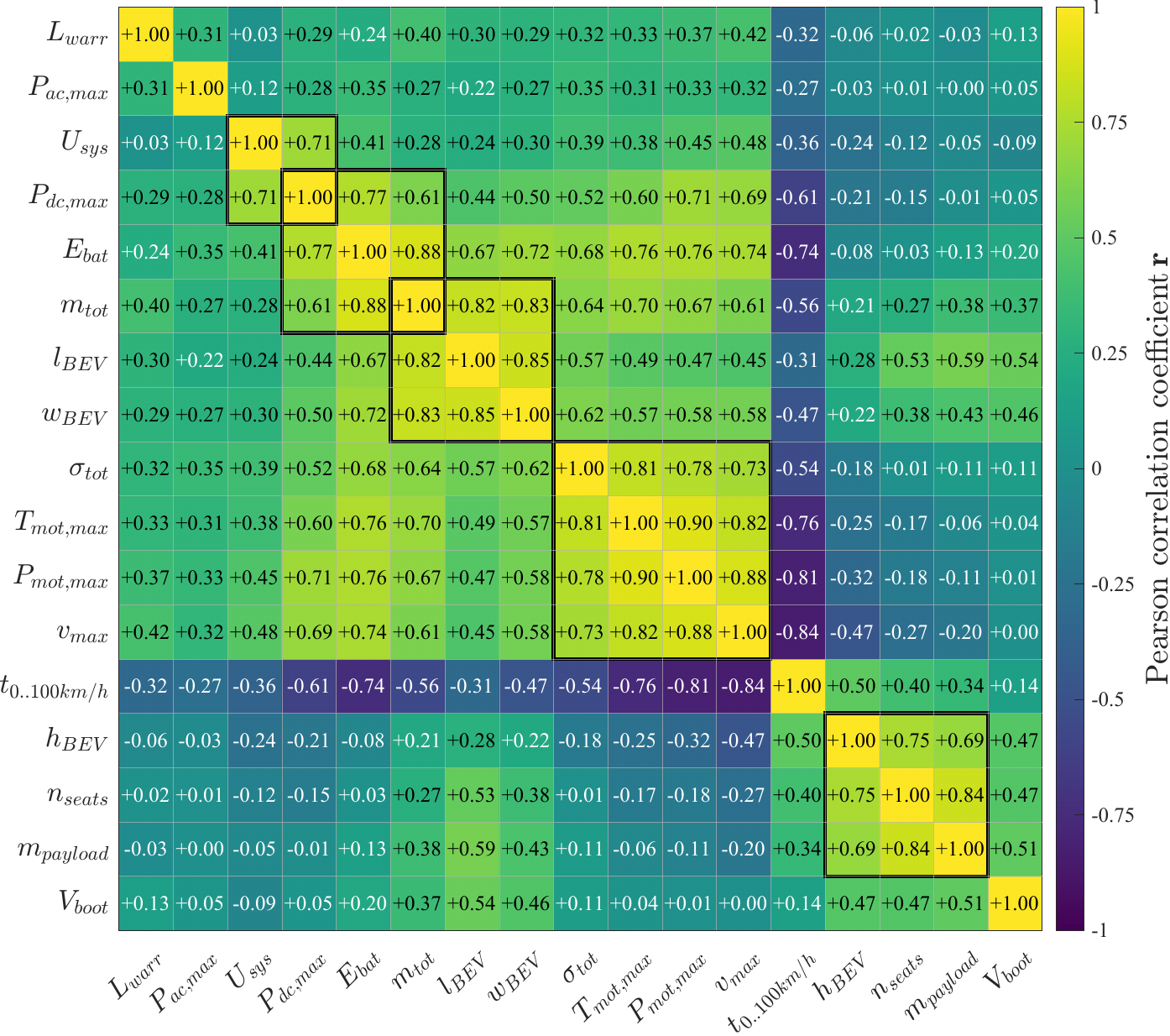}
    \caption{%
    Pearson correlation matrix of several main BEV attributes: 
    warranty distance~$L_{\mathrm{warr}}$,
    maximum AC charging power~$P_{\mathrm{ac,max}}$,
    system voltage~$U_{\mathrm{sys}}$,
    maximum DC charging power~$P_{\mathrm{dc,max}}$,
    battery capacitance~$E_{\mathrm{bat}}$,
    curb weight~$m_{\mathrm{tot}}$,
    vehicle length~$l_{\mathrm{BEV}}$,
    vehicle width~$w_{\mathrm{BEV}}$,
    costs~$\sigma_{\mathrm{tot}}$,
    peak motor torque~$T_{\mathrm{mot,max}}$,
    peak motor power~$P_{\mathrm{mot,max}}$,
    top speed~$v_{\mathrm{max}}$,
    acceleration time~$t_{0\text{--}100~\mathrm{km/h}}$,
    vehicle height~$h_{\mathrm{BEV}}$,
    number of seats~$n_{\mathrm{seats}}$,
    payload~$m_{\mathrm{payload}}$,
    boot volume~$V_{\mathrm{boot}}$. Data filtered based on requirements like variance homogeneity (Breusch-Pagan: $p > 0.05$), normal distribution (Shapiro-Wilk: $p > 0.05$), and outliers ($|z| > 3\,\sigma$).}
    \label{fig:corr-matrix}
\end{figure}

\vspace{-15pt}
\section{Introduction}\label{sec:intro}

Rapid decarbonisation of road transport is essential for meeting the temperature targets set by the Paris Agreement. Legislated fleet-average CO$_2$ limits, exemplified by the stricter European Union standard that applies from model year 2025, push \acp{OEM} towards electric mobility. Global BEV registrations climbed from 2.3\,million units in 2019 to 17\,million units in 2024, lifting market share to approximately 15\,\% \cite{iea2024evoutlook}.

Key performance indicators advanced in parallel. The median \ac{WLTP} driving range rose from 135\,km for 2014 models to 455\,km for 2024 models \cite{doe2024range}. Capital costs per range declined significantly during the same period, reflecting both progress in battery technology and battery scale manufacturing \cite{Link2023_TrendsBatteryCellDesign}. Other trends like the move of battery voltage architectures from 400\,V through 800\,V toward 1000\,V (cf.~\cite{BYD_1000V}) shortens fast-charging sessions to roughly twenty minutes. 

Research attention has lately shifted toward the powertrain itself \cite{rosenberger2024scientific,intro:sierts2024,krueger2024}. Efficiency optimised powertrains boost range and can cut costs simultaneously, as higher drivetrain efficiency enables a battery capacity reduction with equal range, lowering system costs. The traction inverter is central to this optimisation, serving as the DC/AC interface between the battery and motor. Its switching behaviour governs both converter losses and modulation‑induced motor losses. Although the battery dominates the bill of materials, the inverter exerts a disproportionate influence on efficiency, which makes its design pivotal. For instance, replacing \ac{Si} devices with \ac{SiC} wide‑bandgap semiconductors lowers inverter switching losses while preserving adequate blocking margins, thereby improving powertrain efficiency\,\cite{brothers2023sic}.

The three‑phase (3‑Φ) \ac{2L}-\ac{B6} \ac{VSI} is the converter concept most widely deployed in battery‑electric vehicles ($>$99$\%$), because of its structural simplicity, well‑established control principles and robustness against fault events e.g. switching in an active short circuit. Its performance, however, deteriorates significantly at elevated battery voltage architectures: higher‑voltage semiconductor devices must be employed, and the intrinsic \ac{2L}-\ac{B6} switching voltage waveform subjects the machine windings to increased high‑frequency voltage harmonics, thereby increasing modulation‑induced machine losses~\cite{velic2021efficiency}. These modulation-induced losses contribute $\,{}^{\text{1}}\!/\!_{3}$ to the total electric drive losses and can be reduced by $\approx$70\,\% using a 3‑Φ 3L \ac{VSI} at the same switching frequency for a 300\,kW \ac{iPMSM} \cite{sachsECCE2024}.

\subsection{Contributions}

Several recent reviews (cf.~\cite{Poorfakhraei2021review,reimers2019automotive}) {survey inverter topologies and design trends}. However, {they appear to lack quantitative market data and omit optimisation‑based cross‑topology efficiency analyses, so the scale of prospective powertrain efficiency and cost gains remains unclear}.

\begin{figure}[!t]
    \centering
    \includegraphics[width=\linewidth]{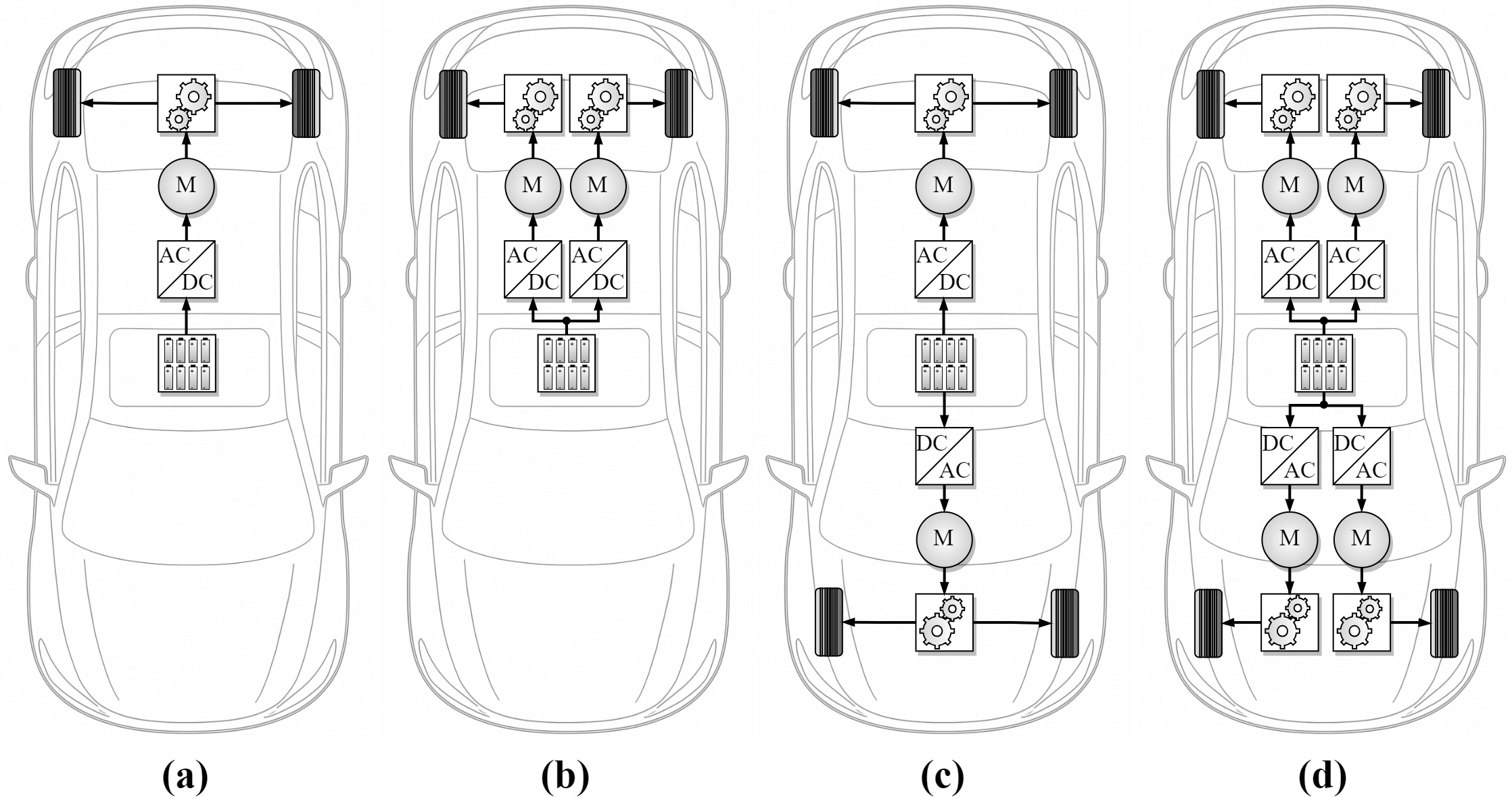}
    \caption{Standard electric powertrain concepts in BEVs: (a) single motor \acp{RWD}, (b) dual-motor \acp{RWD}, (c) dual-motor \acp{AWD} and (d) quad-motor \acp{AWD}.}
    \label{fig:BEVtypes}
    \vspace{-10pt}  
\end{figure}

This study provides insights in fleet trends regarding range,
energy consumption, charging capability and many more \ac{BEV}
properties in Section~\ref{sec:BEVT}. Latest data-based BEV trends are discussed with regards to different powertrain concepts (Fig.~\ref{fig:BEVtypes}),
before an insight in current traction inverter trends is given in Section~\ref{sec:inverter_trends}. The impact of \ac{2L}-\ac{B6} inverters with Si \acf{IGBT}, \ac{2L}-\ac{B6}  inverters with SiC \acf{MOSFET}, and 3L
TNPC and ANPC topologies on drivetrain efficiency
and battery cost savings is assessed in Section IV using a
two-stage simulation framework. Following the methodology
in \cite{velic2021efficiency,sachsECCE2024}, \ac{FEA} motor loss models are coupled with
inverter-loss maps calibrated against laboratory measurements
to obtain cycle-average drivetrain efficiency. Section~\ref{sec:conclusion} summarizes
the findings and concludes future trends.

\subsection{Data set}\label{sec:dataset}

The technical specifications of approximately 1\,000 produced or announced \acp{BEV} models were {screened}. Publicly available sources, including {\ac{OEM}} data sheets, type‑approval {filings}, press releases, technology‑news portals, teardown {notes}, open‑access videos, and peer‑reviewed {articles}, served as inputs. Only vehicle models that are available in Europe or were {formally} confirmed for launch by July~2025 {remained in the set}. Any record missing mandatory fields, namely battery capacity, inverter topology, or a referenced drive‑cycle range, {was excluded}. Facelifted variants adopted the specification of the latest drivetrain inverter and were labelled either \ac{Si} {\ac{IGBT}} or \ac{SiC} {\ac{MOSFET}}. The resulting database spans 72 brands across 12 countries and battery capacities from 16\,kWh to 200\,kWh. A part of the analyzed data within this work is publicly available via \textit{IEEE DataPort}\,\cite{bev_data_2025}.


\begin{figure*}[t!]
    \centering
    \includegraphics[width=1\linewidth]{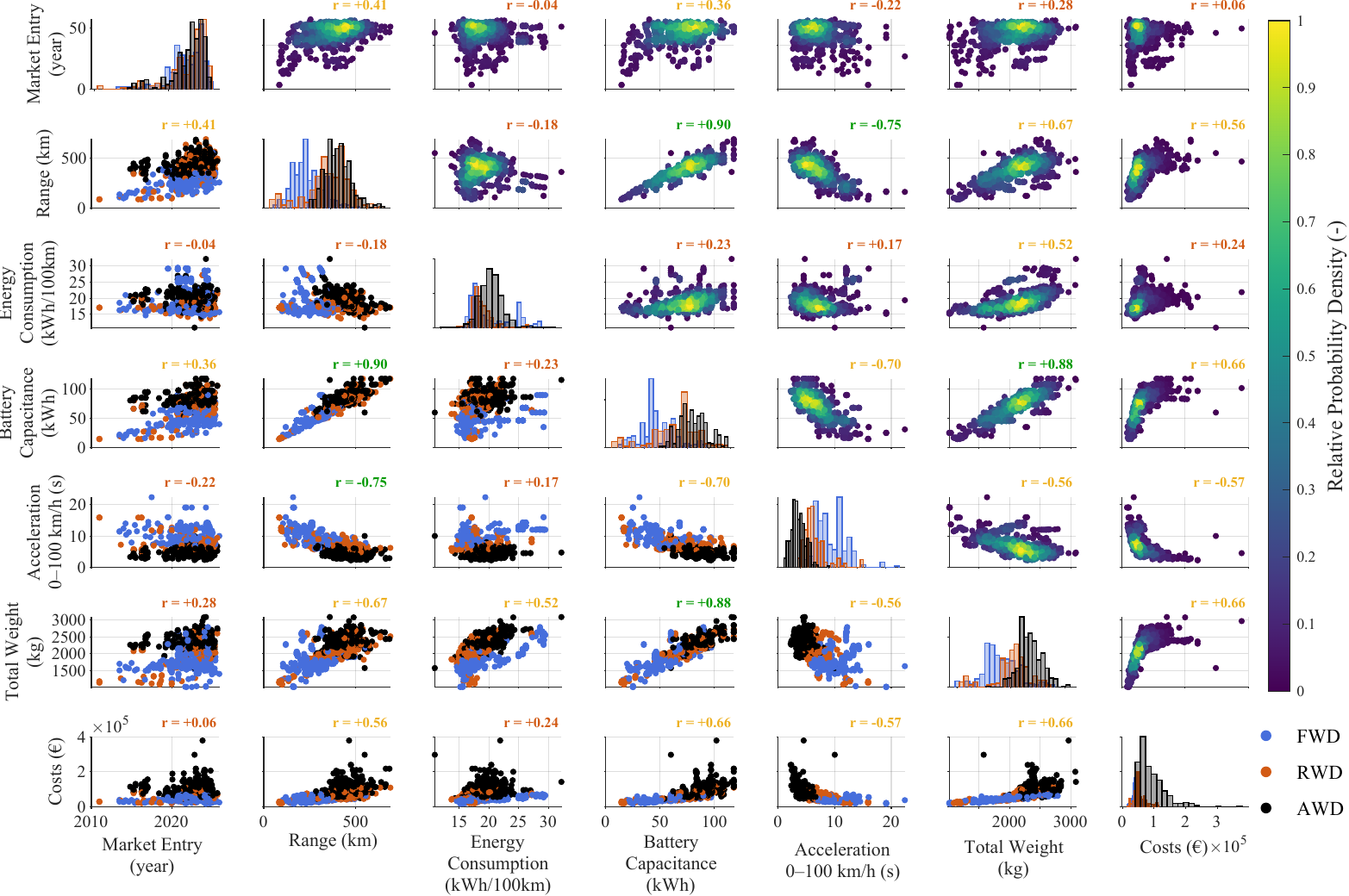}
    \caption{Scatter matrix of key BEV attributes (from 2010 to 2025) illustrating pairwise relationships among vehicle market entry year, driving range, energy consumption, battery capacity, 0–100\,${}^{\text{km}}\!/\!_{\text{h}}$ acceleration time, total weight, and cost. Each cell presents a two-dimensional data cloud for one attribute pair, with the Pearson correlation coefficient $r$ indicated, whereby relative probability density, model clusters, and corresponding histograms of \acf{FWD}, \acf{RWD}, and \acf{AWD} models are shown.}
    \label{fig:scattermatrix1}
    \vspace{-10pt}  
\end{figure*}

\section{BEV Trends}
\label{sec:BEVT}


\begin{table*}[!b]
\vspace{-0.3cm}
\centering
\resizebox{\textwidth}{!}{%
\begin{tabular}{|l|c|c|c|c|c|c|c|c|c|}
\hline
 & $s_{\mathrm{range}}$ (km) & $C_{\mathrm{bat}}$ (${}^{\text{kWh}}\!/\!_{\text{100\,km}}$) & $m_{\mathrm{tot}}$ (kg) & $t_{0\text{--}100}$ (s) & $s_{1\text{--stop}}$ (km) & $E_{\mathrm{bat}}$ (kWh) & $P_{\mathrm{dc,max}}$ (kW) & $m_{\mathrm{tow}}$ (kg) & $V_{\mathrm{boot}}$ (L) \\
\hline
\multicolumn{10}{|c|}{\textbf{2010--2014}}\\\hline
$\mu$      & 134.55 & 17.71 & 1470.91 & 11.74 & 141.45 &  24.55 &  37.18 &  -- & -- \\
$\sigma$   &  87.65 &  1.68 &  336.00 &  3.89 &  87.00 &  19.05 &  22.30 &  -- & -- \\
\hline
\multicolumn{10}{|c|}{\textbf{2015--2019}}\\\hline
$\mu$      & 299.23 & 18.74 & 1890.82 &  7.87 & 293.68 &  56.91 &  63.23 & 283.33 & 501.45 \\
$\sigma$   & 124.13 &  2.83 &  450.29 &  4.22 & 138.27 &  26.48 &  40.84 & 704.52 & 383.73 \\
\hline
\multicolumn{10}{|c|}{\textbf{2020--2025}}\\\hline
$\mu$      & 379.37 & 19.26 & 2112.06 &  7.22 & 414.70 &  72.09 & 116.03 & 911.20 & 514.52 \\
$\sigma$   & 108.38 &  3.11 &  348.74 &  2.94 & 135.97 &  20.16 &  48.81 & 725.05 & 205.47 \\
\hline
\end{tabular}%
}
\caption{Vehicle‐attribute trends in 5‑year intervals: 2010–2014 ($\approx$\,2\,\% of the data), 2015–2019 ($\approx$\,8\,\% of the data), and 2020–2025 ($\approx$\,90\,\% of the data). Each period’s mean ($\mu$) and standard deviation ($\sigma$) are reported for: driving range $s_{\mathrm{range}}$, energy consumption $C_{\mathrm{bat}}$ based on the WLTP (inclusive charging losses), curb weight $m_{\mathrm{tot}}$, $0$–$100$~km~h$^{-1}$ acceleration time $t_{0\text{--}100~\mathrm{km~h^{-1}}}$, one‑stop range $s_{1\text{--stop}}$, battery capacitance~$E_{\mathrm{bat}}$, peak DC fast‑charge power $P_{\mathrm{dc,max}}$, towing capacity $m_{\mathrm{tow}}$, and trunk volume $V_{\mathrm{boot}}$.}
\label{tab:vehicle_data_5yr}
\end{table*}

As shown in Fig.~\ref{fig:corr-matrix} and Fig.~\ref{fig:scattermatrix1}, key performance characteristics of BEVs are strongly interrelated. The scatter matrix in Fig.~\ref{fig:scattermatrix1} reveals the actual interrelation between fundamental variables such as driving range, battery capacity, vehicle mass, acceleration, energy consumption, and cost by showing individual vehicle models in their relative probability density and as members of the groups \acf{FWD}, \acf{RWD} and \acf{AWD} (cf.~Fig.~\ref{fig:BEVtypes}). Fig.~\ref{fig:corr-matrix} on the other hand links further attributes to these key characteristics by solely showing the Pearson correlation coefficient \textit{r} and by highlighting correlation clusters e.g. costs~$\sigma_{\mathrm{tot}}$,
    peak motor torque~$T_{\mathrm{mot,max}}$,
    peak motor power~$P_{\mathrm{mot,max}}$, and
    top speed~$v_{\mathrm{max}}$. This enables a conclusion to be drawn from the detailed attribute distribution shown in Fig.~\ref{fig:scattermatrix1} to the other properties. Contrary to Fig.~\ref{fig:scattermatrix1}, only about ${}^{\text{2}}\!/\!_{\text{3}}$ of all vehicle models have been used to create Fig.~\ref{fig:corr-matrix} as not all detailed information was available for each model, which explains slight differences in shown correlation coefficients. 

\subsection{General analysis (2010--2025)}

According to Fig.~\ref{fig:scattermatrix1}, driving range exhibits a strong positive correlation with battery size (Pearson $r \approx +0.90$), indicating that vehicles equipped with larger battery packs tend to achieve significantly longer ranges. Range is also positively correlated with vehicle weight ($r \approx +0.67$) and cost ($r \approx +0.56$), since long-range models typically carry heavier battery systems and often fall into higher market segments. Conversely, acceleration performance (measured as the 0–100\,${}^{\text{km}}\!/\!_{\text{h}}$ time) correlates negatively with both range ($r \approx -0.75$) and battery capacity ($r \approx -0.70$): modern BEVs offering extended range usually deploy high-power drivetrains, yielding quicker acceleration (lower times) despite their larger mass. This is reflected in the moderate negative correlation between vehicle mass and 0–100\,${}^{\text{km}}\!/\!_{\text{h}}$ time ($r \approx -0.56$), implying that recent heavier models often accelerate faster than older, lighter ones due to substantial improvements in powertrain technology. Vehicles featuring \ac{AWD} drivetrains tend to combine substantial mass, generous battery reserves and higher energy demand, which together push their average costs toward the upper end of the market.

The scatter‑matrix plot hints at a temporal shift in power‑ and energy‑related metrics. Model year correlates moderately with both range and battery capacity ($r \approx +0.41$ and $+0.36$, respectively) and shows a mild inverse link with the 0–100\,${}^{\text{km}}\!/\!_{\text{h}}$ sprint time ($r \approx -0.22$). Taken together, these numbers suggest that between 2010 and 2025 newer \acp{BEV} tended to travel farther on a charge, store more energy, and reach highway speed more quickly. By contrast, model year exhibit no correlation with energy consumption, expressed in ${}^{\text{kWh}}\!/\!_{\text{100\,km}}$ ($r \approx -0.05$). Fleet‑average efficiency therefore appears to have held steady over the period, implying that gains in drivetrain and aerodynamic performance broadly offset the simultaneous rise in vehicle size and weight.


\subsection{Five-year cohort analysis}

Table~\ref{tab:vehicle_data_5yr} summarises how key \ac{BEV} metrics shifted across three consecutive five‑year windows. {Between} the 2010–2014, 2015–2019, and 2020–2025 cohorts, driving range shows a marked rise, {likely driven by} continuing advances in battery chemistry and pack design\,\cite{Link2023_TrendsBatteryCellDesign,saldarini2023battery}. The average {\ac{WLTP} range} almost tripled, growing from roughly 135\,km in 2010–2014 to 379\,km in 2020–2025. During the same span, mean usable battery energy climbed from 24.6\,kWh to 72.1\,kWh (\,$\approx$\,+293\,\%), whereas average BEV mass rose from 1471\,kg to 2112\,kg (\,$\approx$\,+44\,\%). According to Fig.~\ref{fig:corr-matrix}, this weight increase correlates with {greater} vehicle length $l_\text{BEV}$ and width $w_\text{BEV}$. 

Notably, energy capacity grew far faster than either mass or volume, which implies a substantial improvement in battery specific energy \cite{Link2023_TrendsBatteryCellDesign}. Therefore, OEMs now store more energy while accepting only moderate penalties in mass and volume. Average WLTP energy demand rose modestly from 17.7\,${}^{\text{kWh}}\!/\!_{\text{100\,km}}$ to 19.3\,${}^{\text{kWh}}\!/\!_{\text{100\,km}}$. {This delta suggests that powertrain and aerodynamic refinements almost offset the extra mass and higher output of recent \acp{BEV}.} As a reference, the battery-to-wheel consumption is $\approx$\,2\,${}^{\text{kWh}}\!/\!_{\text{100\,km}}$ decreased as it represents the energy consumption of the drivetrain exclusive charging losses. Acceleration capability, by contrast, improved sharply. The mean 0–100\,${}^{\text{km}}\!/\!_{\text{h}}$ sprint dropped from 11.7\,s in 2010–2014 to 7.2\,s in 2020–2025, {nearly a 40\,\% reduction}. {Such progress, achieved despite heavier curb weights, likely reflects advances in motor design and traction‑control enhancements that allow newer models to outpace lighter predecessors.}

\begin{figure}[!t]
  \centering
  \setlength{\tabcolsep}{0pt} 
  \renewcommand{\arraystretch}{1.1}

  \begin{tabular}{cc}
    \includegraphics[width=0.48\linewidth]{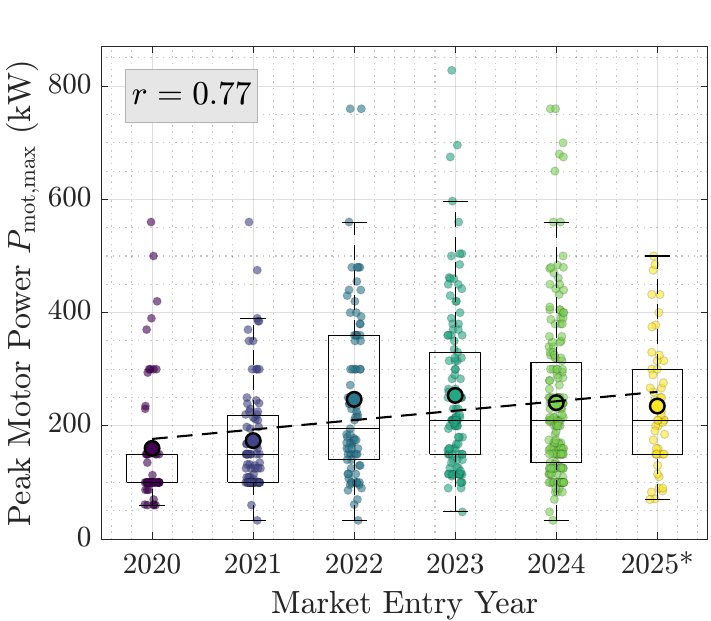} &
    \includegraphics[width=0.48\linewidth]{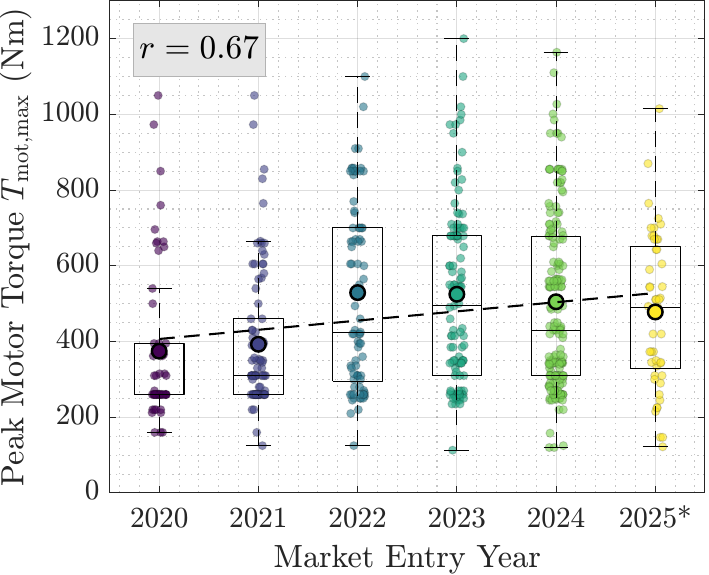} \\
    (a) Motor power $P_{\mathrm{mot,max}}$ &
    (b) Motor torque $T_{\mathrm{mot,max}}$ \\[0.8em]
    \includegraphics[width=0.48\linewidth]{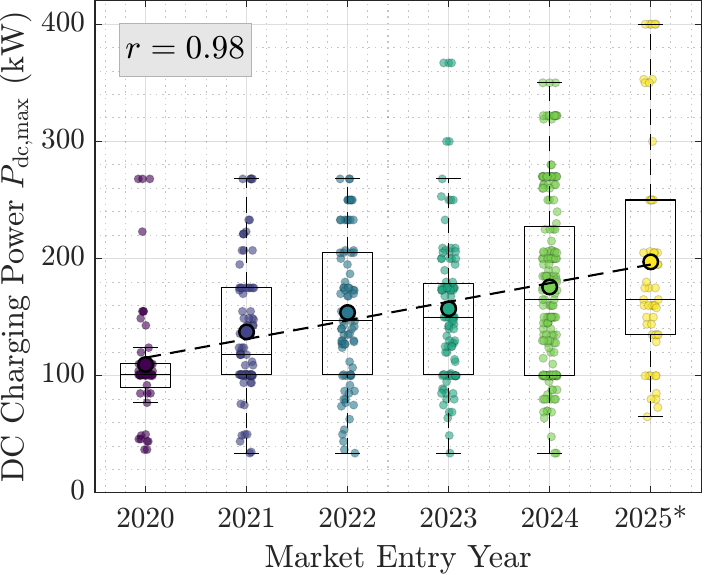} &
    \includegraphics[width=0.48\linewidth]{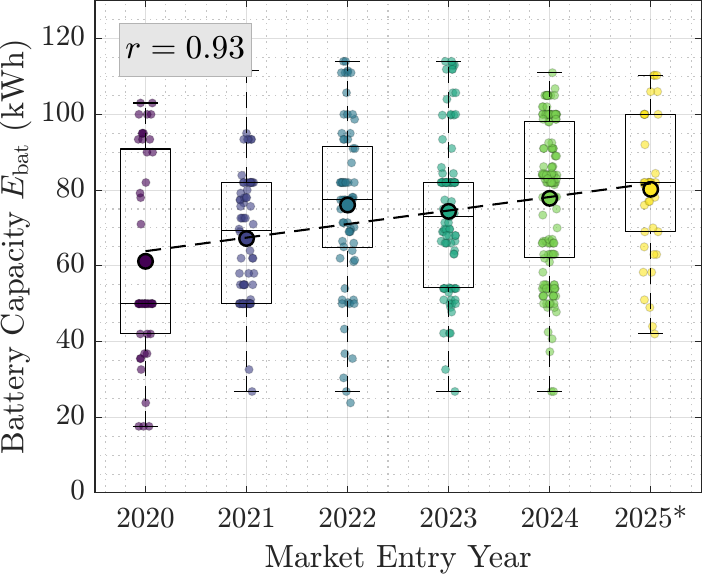} \\
    (c) DC charging power $P_{\mathrm{dc,max}}$ &
    (d) Battery capacity $E_{\mathrm{bat}}$ \\
  \end{tabular}

  \caption{Distribution of key BEV performance
           parameters for models launched between 2020 and 2025, whereby 2025 covers data up to July only.}
  \label{fig:ev_boxplots_all_simple}
  \vspace{-0.5cm}  
\end{figure}

{A further area of rapid improvement concerns fast‑charging and long‑distance travel.} Mean peak DC fast‑charging power climbed from roughly 37\,kW in the early 2010s to about 116\,kW in the early 2020s. {Higher power shortens each stop:} a 2025‑era \ac{BEV} can cover almost three times the distance of a 2012 model after a single fast‑charge session, {greatly enhancing motorway usability}. Over the 15‑year window, \acp{BEV} {appear to have gained} longer range, quicker acceleration, and faster charging while battery specific energy improved. Purchase prices rose only modestly, {whereas extra range emerged as the dominant performance target}.

Fig.~\ref{fig:ev_boxplots_all_simple} summarises four key metrics for recent \acp{BEV}: {(a) peak machine power, (b) peak machine torque, (c) maximum DC‑charging power, and (d) usable battery capacity}. Across all panels an upward trend is apparent, with Pearson coefficients between +0.67 and +0.98, {which suggests that newer model years typically deliver higher performance and quicker charging capabilities}. Battery capacity $E_\text{bat}$ and peak charging power $P_\text{dc,max}$ {have risen especially quickly during the last few years, enabling longer trips with fewer stops and shorter dwell times}. Within the previous years, the fleet‑average peak motor power rose in the direction of $\overline{P}_{\text{mot,max}}= 250\,\text{kW}$. The 2024 distribution spans for the 1$^{\text{st}}$ quartile $P_{\text{mot,max,Q1}} = 150\,\text{kW}$, the 3$^{\text{rd}}$ quartile $P_{\text{mot,max,Q3}}= 300\,\text{kW}$, and a maximum limit without filtered outliers near $P_{\mathrm{mot,max,limit}}= 560\,\text{kW}$. Yet such levels are rarely required on the road \cite{bhaskar2024}. Exemplarily, the \ac{WLTP} cycle tops out at roughly $P_{\text{mot,max,WLTP}} = 50\,\text{kW}$, depending on vehicle mass and drag (cf.~Fig.~\ref{fig:3L_losses}). The gap between installed and used power, as well as maximum full-load and partial-load power losses opens a design space for partial-load optimized inverter concepts, discussed in Section~\ref{sec:inverter}. Looking ahead, manufacturers could hold range steady and channel efficiency gains into smaller, lighter battery packs, which would cut cost and resource demand.


\section{Inverter Technology Trends}\label{sec:inverter_trends}

The traction inverter {largely} determines the drive‑cycle energy efficiency of an electric drivetrain and, by extension, the battery capacity required to meet a given range target. Its influence {extends beyond the converter itself}; the chosen modulation scheme also {excites} additional motor losses \cite{velic2021efficiency,sachs2025modloss}. According to the cost‑ and efficiency‑optimised 300\,kW reference drivetrain analysed in \cite{sachsECCE2024}, electric drive losses break down as follows: {roughly} 10\,\% from inverter switching, 2\,\% from inverter conduction, 55\,\% from fundamental machine losses, and 33\,\% from modulation‑induced machine losses. {The next section discusses latest possibilities to shrink those shares or to hold them steady while trimming hardware cost.}


\begin{figure}[t!]
    \centering
    \includegraphics[width=1\linewidth]{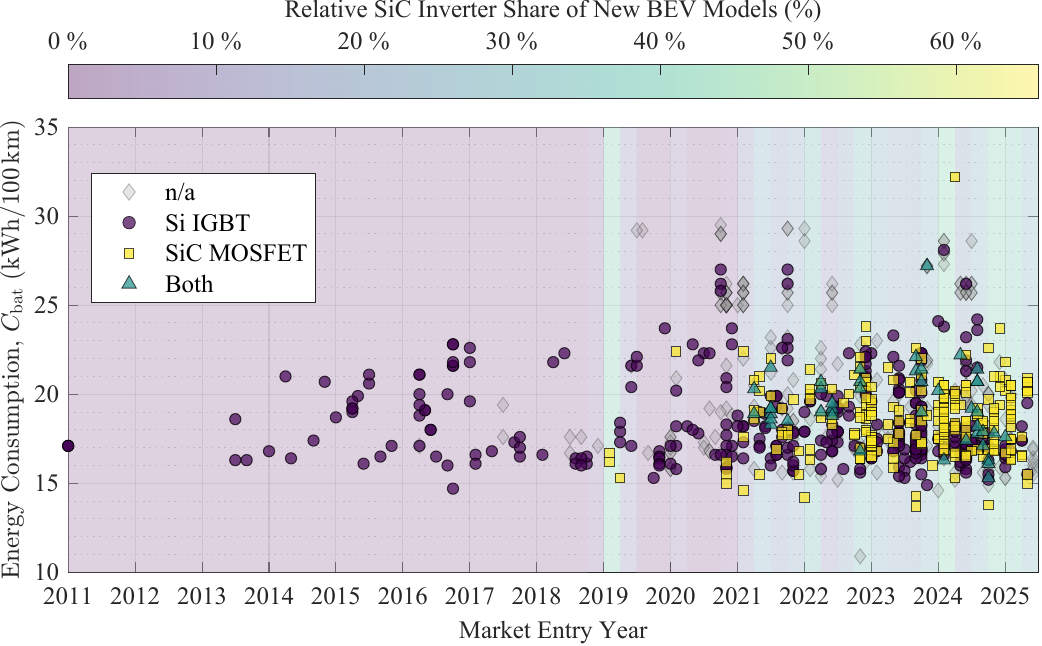}
    \caption{Energy consumption over market entry year of different traction inverter technologies with significant SiC inverter usage starting from 2021.}
    \label{fig:SivsSiC}
    \vspace{-0.5cm}
\end{figure}

\subsection{Wide-bandgap devices and high-voltage architectures}
Since 2021 the move from Si IGBTs to SiC MOSFETs has gained momentum as shown in Fig.~\ref{fig:SivsSiC}. About 50\,\% of new BEVs models either use SiC MOSFET inverters or both SiC and Si e.g. if multiple motors for an AWD are used (cf.~Fig.~\ref{fig:BEVtypes}). Tesla started using SiC MOSFETs for a battery voltage of 400\,V (cf.~\cite{taha2022multiphase}) which were available from February~2019 in Europe as shown in Fig.~\ref{fig:SivsSiC}. {Robert Bosch commenced} volume production of 800\,V \ac{SiC} inverters in 2023. The unit {reportedly} reaches up to 99\,\% efficiency and delivers {about one‑third} higher power density than its \ac{Si} predecessor\,\cite{bosch2023sic}. {Lucid Motors, meanwhile, integrates} a 800\,V \ac{SiC} inverter directly on the traction machine, {packing} 500\,kW into a 74\,kg drive module\,\cite{lucid2022air}.

\subsection{GaN on the horizon}
\begin{figure}[t!]
    \centering
    \includegraphics[width=1\linewidth]{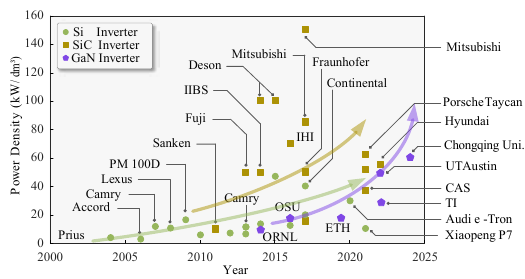}
    \caption{Traction inverter power densities with different wide bandgap semiconductor types over the years with highlighted technology trends, adapted with illustration and data originally from \cite{powerdensity}.}
    \label{fig:GaN_powerdensity}
    \vspace{-10pt}  
\end{figure}

\ac{SiC} devices dominate high‑power series production, {whereas} early \ac{GaN} prototypes hint at the next efficiency jump, as illustrated in Fig.~\ref{fig:GaN_powerdensity}\,\cite{powerdensity}. {The chart plots volumetric power density, expressed in ${}^{\text{kW}}\!/\!_{\text{dm}^3}$, for inverters based on \ac{Si}, \ac{SiC}, and \ac{GaN} switches, covering both laboratory demonstrators and market‑ready units from industry and research groups}. Several points correspond to alternative topologies such as 3-Φ 3L \ac{ANPC} converters. 

Exemplarily, Cambridge GaN Devices {exhibited} a \(3\text{L}\) \ac{ANPC} inverter, whereby its GaN transistors are designed for 800\,V traction applications\,\cite{cgd2025pcim}. Although still pre‑series, these results {may suggest} that \ac{GaN} technology could soon rival, or even surpass, \ac{SiC} in high‑power automotive drives.

Although \ac{GaN} devices {permit} higher switching frequencies with only modest converter loss\,\cite{xu2021impact,10458519}, {the attendant rise in parasitic motor losses often goes unnoticed}. Raising \(f_\text{sw}\) reduces the familiar modulation‑induced loss components in copper, iron, and magnets, 
\(
(P_\text{cu,h}+P_\text{iron,h}+P_\text{mag,h})\propto f_\text{sw}^{0.5\,..\,0.7},
\)
yet it also {magnifies} losses linked to inter‑winding and winding‑to‑stator capacitances, 
\(
P_\text{capa,h}\propto f_\text{sw},
\)
as reported in \cite{JanMatrix2025,sachs2025modloss}. Therefore, a holistic efficiency comparison of inverters should incorporate the operation impact on the electric motor.

\subsection{Thermal packaging and power-density push}

Higher switching frequencies and current densities {appear to make} thermal management the next bottleneck. BorgWarner’s most recent \ac{DSC} 800\,V \ac{SiC} power module, extracts heat from both die surfaces and {is said to enable} about 30\,\% more current or a smaller heatsink at the same output\,\cite{borgwarner2025dsc}. {Component design is progressing in parallel.} Infineon’s trench super‑junction \ac{SiC} \ac{MOSFET} lowers the figure‑of‑merit \(R_{\mathrm{DS(on)}}\!\times\!A\) by roughly 40\,\%, which could permit nearly 25\,\% higher phase current without enlarging the package\,\cite{infineon2025sj}. Additional benefits of \ac{DSC} packaging {are examined in detail} in \cite{liu2022comprehensive}.

Cost {now competes with} efficiency as the dominant design driver, as {summarised} in Section~\ref{sec:BEVT}.  
Tesla’s 2023 Investor‑Day slide {forecast} a 75\,\% reduction in \ac{SiC} die area for its next drive unit, {placing} cost near 1\,000\,\$ with no performance loss (cf.~\cite{tesla2023investor}).  
{Tier‑1 suppliers have answered with modular concepts.} ZF’s SELECT platform (2025) {accepts} 400\,V \ac{Si}, 400/800\,V \ac{SiC}, or future \ac{GaN} stages in the same mechanical envelope and could cut \ac{OEM} development time by up to 50\,\%\,\cite{zf2025select}.  
Robert Bosch {pursues} a similar strategy, {offering} its \ac{SiC} inverter as either a stand‑alone unit or an element of a 3‑in‑1 e‑axle\,\cite{bosch2023sic}.

\subsection{Multi-functional subsystems}\label{sec:multif}

In 2024, Infineon {presented} an inverter concept built around {so‑called} hybrid power modules\,\cite{bauer2024optimization}. {Each module parallels} \ac{Si} \acp{IGBT} with \ac{SiC} \acp{MOSFET}, while dedicated gate‑driver strategies are used to maintain high efficiency under part‑load conditions. {By reserving more expensive \ac{SiC} switches for operating points where their low switching loss matters most, the design seeks to balance performance with component expense rather than rely solely on \ac{SiC} transistors.}

\begin{figure}[t!]
    \centering
    \includegraphics[width=1\linewidth]{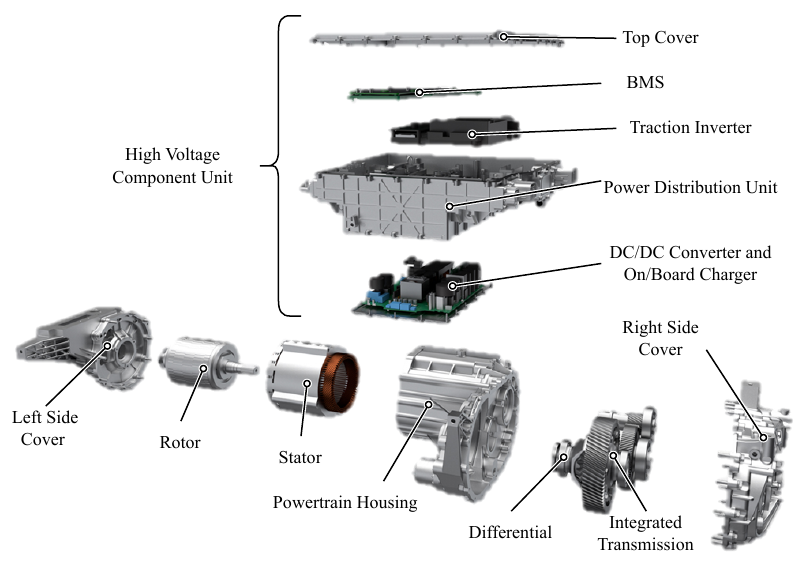}
    \caption{Exploded view of a BYD integrated 8-in-1 powertrain which combines several power electronic subsystems in a single unit, adapted from \cite{byd}.}
    \label{fig:BYD}
    \vspace{-10pt}  
\end{figure}

Building multifunctional subsystems ranks among the most effective ways to trim cost. {For instance,} BYD became the first manufacturer to mass‑produce an 8‑in‑1 drive unit (see Fig.~\ref{fig:BYD}). The assembly {packages} the traction machine, two‑stage gearbox, inverter, DC/DC converter, \ac{OBC}, \ac{PDU}, vehicle‑control electronics, and the \ac{BMS} inside a single sealed housing. Co‑locating these elements removes separate enclosures, high‑voltage (HV) cabling, mounting brackets, and stand‑alone coolant circuits, {which in turn} lowers material cost, cuts assembly steps, and reduces connector‑related failure risk\,\cite{electronics_cooling_2022}. {Thermal behaviour improves as well}: a shared coolant jacket and cold‑plate distribute heat more evenly, permitting either higher silicon junction temperatures or smaller heat exchangers at unchanged lifetime targets. Shorter HV busbars also trim conduction loss and electromagnetic interference, while the compact package frees under‑body volume for additional battery cells or passenger space.

{Hyundai and Kia apply} a functional strategy on the E‑GMP platform. During 400\,V DC fast charging the rear traction machine and its \ac{SiC} inverter {are used as} a boost converter, lifting the bus voltage to 800\,V without an extra high‑power DC/DC stage\,\cite{hyundai2021multi}. {Re‑using} existing hardware sidesteps a dedicated step‑up converter, trims cost, and leaves coolant routing unchanged. {The same inverter topology lets the motor operate more efficiently in the base‑speed region} (cf.~\cite{sachsECCE2024}). Earlier groundwork came from the Renault Zoe: its Caméléon concept re‑purposes the traction inverter as a three‑phase active rectifier and employs the stator windings as boost inductors, enabling up to 43\,kW AC charging with no stand‑alone \ac{OBC} but leads to increased charging losses \cite{subotic2016integrated}. {Comparable integration schemes now appear in other electric‑mobility segments, including high‑power scooters\,\cite{Escooter}.}

{Pulsetrain} introduced a system that combines the high-voltage pack with multiple power-electronic functions: BMS, OBC, DC/DC stage and a MLI inside one compact in-battery module. {According to the supplier, the arrangement may deliver roughly twice the charging speed, trim drivetrain mass by about 40\,\%, and extend cell life by up to 80\,\%.} The concept has yet to enter series production\,\cite{pulsetrain2025}. Comparable approaches {appear under study} at several \acp{OEM}, including Porsche\,\cite{porsche_ACbat} and Mercedes‑Benz\,\cite{mercedes_vehicle_2025}.

\subsection{Software optimization}\label{sec:SW}

{Drive‑cycle efficiency can be increased by tuning modulation and control parameters, for example through operating‑point‑dependent discontinuous \ac{PWM}, adaptive switching frequency, or an alternative space‑vector sequence.} {Such choices influence inverter switching loss, time‑harmonic motor loss, and common‑mode voltage, as reported in} \cite{krueger2024,sachsECCE2024,velic2021efficiency,sachs2025modloss,bauer2024optimization,kolar2006analytical,pwm_analysis}.

 \section{Efficiency/Cost Analysis}\label{sec:inverter}
\begin{figure}[t!]
    \centering
    \includegraphics[width=\linewidth]{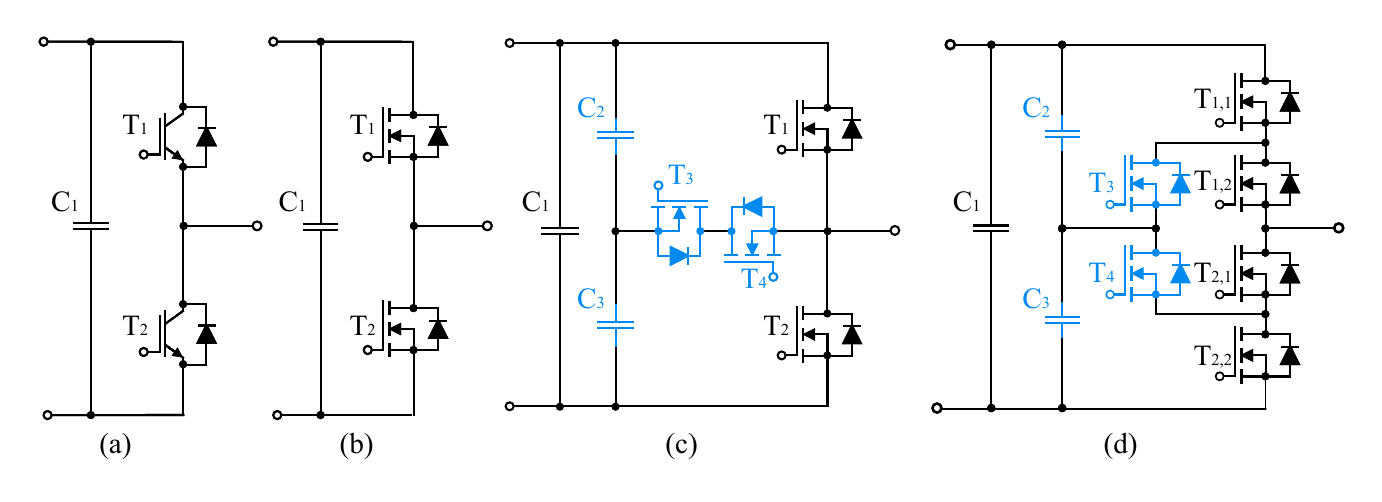}
    \caption{Traction inverter topologies under investigation: (a) 3-Φ \ac{2L}-\ac{B6} converter with Si IGBTs, (b) 3-Φ \ac{2L}-\ac{B6} converter with SiC MOSFETs, (c) 3-Φ \ac{2L}/3L-TNPC with SiC MOSFETs, and (d) 3-Φ \ac{2L}/3L-ANPC with SiC MOSFETs. Components used solely for partial-load operation (3L operation) are highlighted in blue: C$_2$, C$_3$, T$_2$, and T$_3$.}
    \label{fig:topologies}
    \vspace{-13pt}  
\end{figure}

Three inverter topologies are evaluated with respect to energy efficiency and potential cost savings (cf.~Fig.~\ref{fig:topologies}): a conventional 3-Φ \ac{2L}-\ac{B6} inverter equipped with (a) SiC and (b) Si switches, a partial-load optimised 3-Φ 3L-TNPC (c), and a 3-Φ 3L-ANPC based on SiC devices (d) likewise realized with SiC technology. The term \emph{partial-load optimised} refers to the design methodology first introduced in \cite{sachsECCE2024} and exemplarily implemented in a DeepDrive/FEV prototype \cite{rosen2025elektrische}. In this approach, maximum torque and inverter power are achieved even when only a subset of the semiconductor switches is active.
The left partial‑load switches in a topology, i.e. T$_3$ and T$_4$ highlighted in blue in Fig.~\ref{fig:topologies} enable a more efficient 3L operation. Therefore, they can be optimised solely for specific, so-called partial-load operation points. This is relevant, considering that actual occurring operation points usually tap only a small fraction of the peak power capability of current motor drives and high power operation points occur very infrequently \cite{sachsECCE2024,intro:sierts2024,bhaskar2024}.

{The design process begins by selecting the devices that govern full‑power operation, for example T$_1$ and T$_2$ in a \ac{TNPC} inverter (cf.~Fig.~\ref{fig:topologies}).} Once those switches satisfy the peak‑load requirement, the partial‑load devices T$_3$ and T$_4$ are {dimensioned so the converter still fulfills the design goals: In our case it is capable of delivering 80\,\% of the maximum‑torque‑per‑volt output in 3L mode at ${}^{\text{2}}\!/\!_{\text{3}}$ of the maximum motor speed}. 

{Two limits define the admissible operating region:} the device junction temperature may not exceed $\vartheta_\text{max}=175\,^\circ\text{C}$, and the ripple on the 800\,V DC-link must stay within $\Delta U \leq \pm5\,\%$ while the inverter is operated with 10\,kHz SV\ac{PWM}. Chip area $A_\text{chip}$ as main cost driver is increased only as far as needed to meet those constraints, a step‑wise strategy that can lower overall power‑train loss while adding minimal cost\,\cite{sachsECCE2024}.

\subsection{Model assumptions}

Semiconductor switch models originate from data-sheet information complemented by laboratory measurements. 4$^\text{th}$-generation SiC MOSFETs from ROHM (\textit{SCT4018KR} and \textit{SCT4013DR}) are employed, as specified in~\cite{ROHM_datasheet}. Si IGBTs devices are represented by 7$^\text{th}$-generation Trenchstop IGBTs (\textit{IKQ120N120CS7}) from Infineon~\cite{Infineon_IKQ120N120CS7}.  
The electrical machine is based on a 300\,kW \ac{iPMSM} FEA model. 

\subsection{Energy losses and consumption forecast}

The total drivetrain power losses \(P_{\text{tot}}\) comprise three components: the inverter semiconductor losses (switching and conduction losses), modulation-induced motor losses, and fundamental-frequency motor losses,

\begin{equation}\label{eq:Ptot}
    P_{\text{tot}} = \sum_{i=1}^{m} P_{\text{switch},i} + P_{\text{mot,h}} + P_{\text{mot,f}}.
\end{equation}

\begin{figure}[t!]
  \centering
  \begin{tikzpicture}
    \node[inner sep=0pt] (main) {%
      \includegraphics[width=\linewidth]%
        {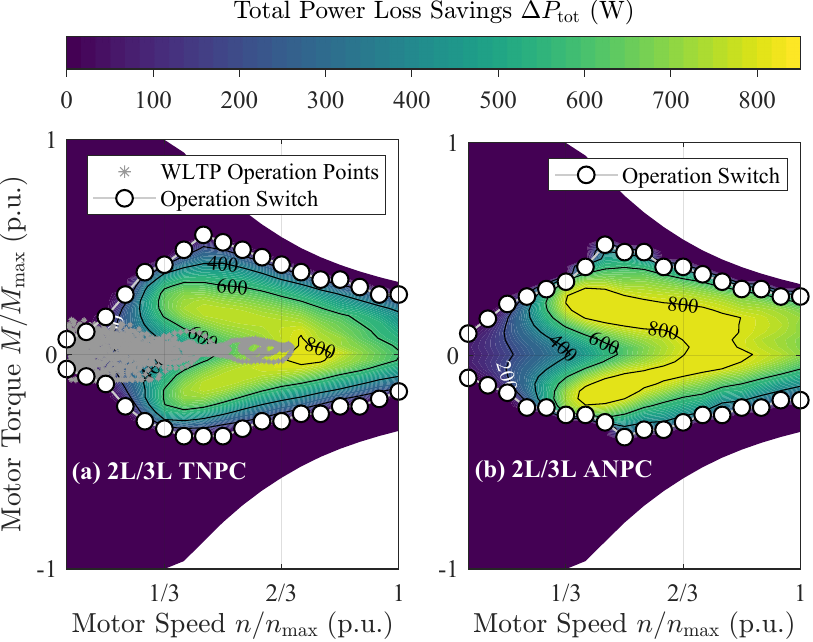}};

    \def\insetwTNPC{0.13\linewidth}   
    \def\xoffTNPC{3.3cm}              
    \def\yoffTNPC{0.65cm}             

    \def\insetwANPC{0.13\linewidth}
    \def\xoffANPC{0.07cm}              
    \def\yoffANPC{0.75cm}

    \node[overlay, anchor=south east, inner sep=0pt]
         at ([xshift=-\xoffTNPC-\insetwTNPC,yshift=\yoffTNPC]main.south east)
         {\includegraphics[width=\insetwTNPC]%
            {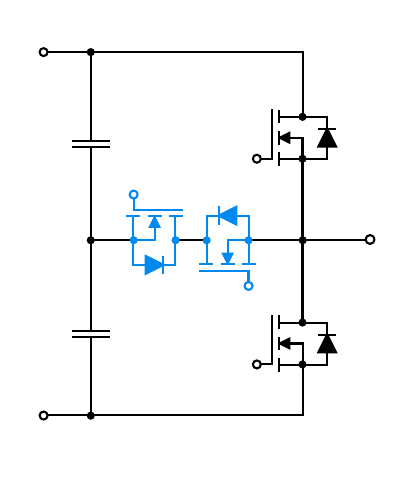}};

    \node[overlay, anchor=south east, inner sep=0pt]
         at ([xshift=-\xoffANPC,yshift=\yoffANPC]main.south east)
         {\includegraphics[width=\insetwANPC]%
            {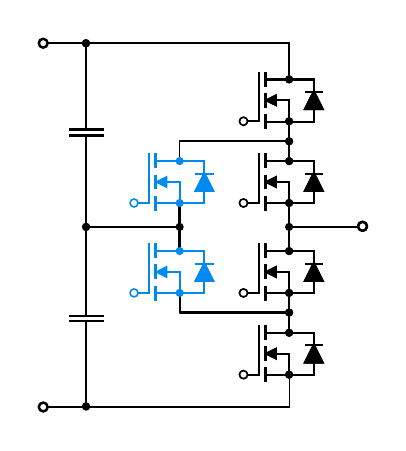}};
  \end{tikzpicture}

  \caption{Power loss savings $\Delta P_\text{tot}=P_\text{tot,2L}-P_\text{tot,3L}$ of (a) \ac{2L} and \ac{3L} operation for a 3‑Φ \ac{2L}/3L‑TNPC SiC MOSFET\,VSI on the left (chip area $A_\text{chip}$ increase of 30\,\% versus a 3‑Φ \ac{2L}‑B6\,VSI, cf.~\cite{sachsECCE2024}) and (b) a 3‑Φ \ac{2L}/3L‑ANPC SiC MOSFET\,VSI on the right ($A_\text{chip}$ increase of 69\,\% versus a 3‑Φ \ac{2L}‑B6\,VSI).}
  \label{fig:3L_losses}
  \vspace{-10pt}
\end{figure}

Thereby, \(P_{\text{switch},i}\) denotes the switching losses of device \(i..m\), \(P_{\text{mot,h}}\) the harmonic (modulation-related) motor losses, and \(P_{\text{mot,f}}\) the fundamental motor losses. A detailed calculation procedure for both inverter and machine losses is provided in~\cite{sachsECCE2024,sachs2025modloss}. The partial-load operation limits (due to introduced constrains) and power loss savings of both TNPC and ANPC inverters are shown in Fig.~\ref{fig:3L_losses}.

Inverter and motor energy losses over the WLTP driving cycle are obtained by summing up the total system losses \(P_{\text{tot}}\) during the WLTP with a normalisation to 100\,km range:

\begin{equation}\label{eq:deltaE}
\Delta E = \left[ \frac{100 \text{\,km}}{s_{\text{WLTP}}} \right] \times \int_{0}^{T_{\text{WLTP}}} P_{\text{tot}} \, \text{d}t,
\end{equation}

yielding \(\Delta E\) in ${}^{\text{kWh}}\!/\!_{\text{100\,km}}$. Lower values indicate improved drivetrain efficiency and a reduced battery capacity. The corresponding cost impact is estimated by

\vspace{-0.5cm}

\begin{equation}\label{eq:costs}
    \Delta\sigma_{\text{tot\,[\euro]}} = \Delta E_{\text{\,[kWh/100\,km]}} \, \times \, s_{\text{[100\,km]}} \, \times \, \sigma'_{\text{bat\,[\euro/kWh]}},
\end{equation}

where \(\sigma'_{\text{bat}}\) = ${70\,}^{\text{\euro}}\!/\!_{\text{kWh}}$ denotes the specific battery cost as a BEV battery cost forecast for 2030 from Goldman Sachs \cite{goldman2024lower}.


\begin{table}[h!]
\centering
\caption{Energy consumption and cost saving forecasts for 2030 of different inverter topologies compared to a \ac{2L}-\ac{B6} SiC VSI over various driving ranges.}
\label{tab:inverter_savings}
\begin{tabular}{lccc}
\toprule
\textbf{Topology} & {Range [km]} & \(\Delta E\) [${}^{\text{kWh}}\!/\!_{\text{100\,km}}$] & \(\Delta \sigma\) [\euro] \\
\midrule
Si \ac{2L}-\ac{B6} \newline
    & 300 &  1.351 &  +94.58 \\
    & 500 &  2.252 & +157.64 \\
    & 700 &  3.153 & +220.70 \\
\midrule
SiC 3L-TNPC  
   & 300 & -2.009 & -140.62 \\
   (+30\,\% $A_\text{chip}$)  & 500 & -3.348 & -234.36 \\
    & 700 & -4.687 & -328.10 \\
\midrule
SiC 3L-ANPC \newline 
    & 300 & -2.339 & -163.72 \\
 (+69\,\% $A_\text{chip}$)    & 500 & -3.898 & -272.86 \\
    & 700 & -5.457 & -382.00 \\
\bottomrule
\end{tabular}
\end{table}

The table summarizes the energy consumption \(\Delta E\) and corresponding cost \(\Delta \sigma\) in \euro~for each inverter topology over 300~km, 500~km, and 700~km of driving range, relative to the \ac{B6} SiC baseline.  Negative values represent improvements in energy efficiency and lower cost, while positive values indicate worse performance. The relative increase in required SiC chip area compared to a \ac{2L}-\ac{B6} inverter is shown on the left. Additional semiconductor costs are not implemented in $\Delta \sigma$, but represent the main cost driver for SiC inverters, leading to a far more cost-effective 2L/3L-TNPC inverter compared with the 2L/3L-ANPC concept.

\subsection{Further partial-load MLI improvements}

In addition to the hybrid switch modules from Infineon \cite{bauer2024optimization}, MLIs are   suitable for hybridisation too. Exemplarily, the partial-load switches of the ANPC can be designed as SiC and full load switches as Si IGBTs \cite{woldegiorgis2021high}. This combines the efficiency increase through SiC MOSFETs with cost reduction of a inverter capable of supplying full load operation.  

Operating 3L inverter with serial capacitors typically increases the required DC-link capacitance. The design methodology proposed in~\cite{sachsASIA2025} mitigates this penalty by introducing a novel split DC-link design (cf.~Fig.~\ref{fig:topologies}) with partial-load optimisable capacitors C$_2$, and C$_3$ to achieve a 34\,\% reduction in DC-link storage energy and a 71\,\% reduction in total capacitance for a \ac{2L}/3L-TNPC compared with a standard 3L-TNPC topology. Further improvements can be achieved using optimised pulse pattern as introduced in \cite{hepp2024} to cut down the actual voltage ripples of the DC-link capacitors and decrease the required capacitance if the thermal system mitigates the occurring current design influence \cite{kolar2006analytical,sachsASIA2025}.  


\section{Conclusion}\label{sec:conclusion}

The analysis of more than 1\,000 European BEV models introduced between 2010 and 2025 shows that the median WLTP range has tripled. Yet, average BEV WLTP energy consumption has remained almost unchanged, indicating that efficiency increases have offset the mass and size growth of modern BEVs. Co-simulation further indicate that a partial-load optimised \ac{2L}/3L TNPC SiC inverter reduces WLTP drivetrain losses by 0.67\,${}^{\text{kWh}}\!/\!_{\text{100\,km}}$ relative to a state-of-the-art \ac{2L}-\ac{B6} SiC inverter while requiring only 30\,\% additional chip area, corresponding to $\approx$\,234\,\euro~battery-pack savings for a 500\,km vehicle range. Partial-load MLIs therefore represent one of the few remaining levers capable of simultaneously lowering energy consumption and total system cost in future BEV powertrains.


\acrodef{FEA}{finite element analysis}
  \acrodefplural{FEA}[FEAs]{finite element analyses}
  
\acrodef{BEV}{battery electric vehicle}
  \acrodefplural{BEV}[BEVs]{Battery electric vehicles} %

\acrodef{BMS}{battery‑management system}%
  \acrodefplural{BMS}[BMSs]{battery‑management systems}%

\acrodef{WLTP}{worldwide harmonized light vehicles test procedure}%

\acrodef{OEM}{original equipment manufacturer}%
  \acrodefplural{OEM}[OEMs]{original equipment manufacturers}%
\acrodef{iPMSM}{interior permanent‑magnet synchronous motor}%
  \acrodefplural{iPMSM}[iPMSMs]{interior permanent‑magnet synchronous motors}%


\acrodef{WBG}{wide-bandgap}
\acrodef{GaN}{Gallium Nitride}
\acrodef{SiC}{silicon carbide}
\acrodef{Si}{silicon}
\acrodef{IGBT}{insulated‑gate bipolar transistor}%
\acrodef{MOSFET}{metal‑oxide semiconductor field‑effect transistor}%


\acrodef{VSI}{voltage-source inverter}
  \acrodefplural{VSI}[VSIs]{voltage-source inverters}%

\acrodef{MLI}{multi-level inverter}
\acrodef{2L}{two-level}
\acrodef{3L}{three-level}
\acrodef{5L}{five-level}
\acrodef{3-Φ)}{three-phase}

\acrodef{TNPC}{T-type neutral point clamped}
\acrodef{ANPC}{active neutral point clamped}
\acrodef{B6}{six halfbridge}


\acrodef{PWM}{pulse‑width modulation}%
\acrodef{RMS}{root‑mean‑square}%


\acrodef{AWD}{all‑wheel drive}%
  \acrodefplural{AWD}[AWDs]{all‑wheel drives}
\acrodef{FWD}{front‑wheel drive}%
  \acrodefplural{FWD}[FWDs]{front‑wheel drives}%
\acrodef{RWD}{rear‑wheel drive}%
  \acrodefplural{RWD}[RWDs]{rear‑wheel drives}%

\acrodef{BMS}{battery‑management system}%
\acrodef{DSC}{double‑sided‑cooled}%
\acrodef{OBC}{on‑board charger}%
\acrodef{PDU}{power‑distribution unit}%

\vspace{-0.7cm}
{\setstretch{1}\vspace{\baselineskip}
\bibliographystyle{IEEEtran}
\bibliography{00_main.bib}
}



\end{document}